# Narrow and contrast resonance of increased absorption in Λ-system observed in Rb cell with buffer gas

A. Sargsyan[1], A. Papoyan[1], A. Sarkisyan[1], Yu. Malakyan[1], G. Grigoryan[1], D. Sarkisyan[1*]

[1]*Institute for Physical Research, NAS of Armenia, Ashtarak, 0203, Armenia*

Y. Pashayan-Leroy[2], C. Leroy[2]

[2]*Institut Carnot de Bourgogne, Université de Bourgogne, B.P.47870, 21078 Dijon cedex, France*

**Abstract**

We report observation of a narrow (sub-natural) and high-contrast resonance of increased absorption ("bright" resonance) in Rb cell with Ne buffer gas under previously unexplored experimental conditions for coupling and probe radiation configuration. The coupling laser stabilized frequency is detuned by ~ 3 GHz from $5S_{1/2}$, $F_g=3 \rightarrow 5P_{3/2}$, $F_e=2,3,4$ transitions, while the probe laser frequency is scanned across these transitions. We believe the bright resonance formation, occurring when the probe laser frequency is blue-shifted from the coupling frequency by a value of the ground state hyperfine splitting, is caused predominantly by a 2-photon absorption of the probe radiation $5S_{1/2}$, $F_g=2 \rightarrow 5S_{1/2}$, $F_g=3$ with $5P_{3/2}$ as an intermediate state. We also report and interpret splitting of the bright resonance into 6 well resolved and contrast components in moderate magnetic fields ($B \sim 10 \div 250$ G).

## I. INTRODUCTION

There has been considerable interest in recent years for the fascinating properties of Coherent Population Trapping (CPT) and the related Electromagnetically-Induced Transparency (EIT) arising due to destructive interference of atomic transitions, as well as of Electromagnetically-Induced Absorption (EIA) occurring due to constructive interference. These phenomena attract much attention because of their significance for various applications in metrology, magnetometry and fundamental investigations (see the papers and reviews [1-10]). The EIT and CPT resonances can occur in a Λ-system with two long-living ground states and one excited state coupled by two laser fields: one laser frequency is resonant with the atomic transition, while the frequency of the second (probe) one is scanned across the upper atomic transitions, forming so–called narrow "dark" resonance inside the Doppler–broadened transmission spectrum of the probe laser. The EIA can occur in a V-type system [6]: in this case the ground level should be degenerated and the quantum number of the upper level $F_e$ should be larger than that of the ground level $F_g$. To observe EIA, one laser frequency should be resonant with the atomic transition, while the second one should be scanned across the atomic transitions, forming so called narrow "bright" resonance inside the Doppler–broadened transmission spectrum of the probe laser.

In order to achieve narrow EIT or EIA resonance, two lasers should be coherently coupled. For this purpose several modulation techniques could be implemented [7]. However, there are some cases when a large frequency region of the probe laser frequency tuning is needed, and for this case the use of two different (independent) lasers serving as coupling and probe radiation sources is the only solution.

Here we report a new scheme, which allows one to form a "bright" resonance inside the Doppler-broadened transmission spectrum of the probe laser using a cell filled with natural Rb and buffer gas neon. In spite of different formation conditions, this narrow "bright" resonance inside the probe radiation transmission spectrum behaves similar to "genuine" EIA resonance observable in a V-system [6]. We suppose the physical mechanism responsible for this resonance formation is a two-photon absorption of the probe radiation $5S_{1/2}$, $F_g=2 \rightarrow 5S_{1/2}$, $F_g=3$ through the intermediate state $5P_{3/2}$. The lineshape of this resonance is also affected by some competing process taking place in the same experimental conditions.

## II. EXPERIMENTAL ARRANGEMENT AND TECHNIQUE

The experimental arrangement is sketched in Fig.1. The beams ($\varnothing$ 3 mm) of two separate single-frequency extended cavity diode lasers (ECDL) with ~1 MHz spectral linewidth, which serve as coupling and probe beams with $\lambda \approx 780$ nm, after passage through Faraday isolators (*1*) are well superposed, with the help of the first Glan prism, and directed onto the 8-mm long sealed-off T-shaped all-sapphire cell (ASC) (*4*) [11] filled with natural mixture of $^{85}$Rb and $^{87}$Rb and 6 Torr of neon buffer gas. A special oven made of non-magnetic materials has been used to provide needed thermal regime of operation: the temperature was $T_W \sim 100^oC$ at the ASC windows and $T_{SA} \sim 80^oC$ at the side-arm. The power of coupling and probe lasers having mutually perpendicular linear polarizations was varied in the range of 0.01 – 15 mW. The second Glan prism is used to cut the coupling laser beam after the passage through the cell. The cell was placed inside the three mutually perpendicular pairs of Helmholtz coils (*2*) providing a possibility to cancel the laboratory magnetic field as well as to apply a homogeneous magnetic field of needed strength and direction. The optical radiations were recorded by the photodiodes with operation amplifiers (*3*) followed by a 4-channel digital storage oscilloscope Tektronix TDS 2014B.

A modified DAVLL method realized in a separate nano-cell with the thickness $L = \lambda/2$ is used for stabilization of the coupling laser frequency. The idea is as follows: in the DAVLL method, an ordinary cell with the thickness of several centimeters is used to form an error signal [12]. The use of a nano-cell with the thickness $L = \lambda/2$ allows one to obtain sub-Doppler absorption spectrum (see [13,14]), which is ~ 4 times narrower than that obtained with an ordinary cm-size cell. Thus, the slope of the error signal and hence efficiency of frequency lock is much better than that reported in [12]. A longitudinal magnetic field is formed with the help of a permanent ring magnet marked PM in Fig.1, which has a ∅ 2 mm hole for the laser beam propagation. The ring magnet was mounted on the micrometric translation stage for longitudinal displacement needed to control the $B$-field. Variation of direction and strength of an external magnetic field modifies the error signal and causes an overall frequency shift. In spite of strong inhomogeneity of a $B$-field produced by a ring magnet, the variation of $B$ inside atomic vapor column is only a few µT due to small thickness of the nano-cell (780 nm) [15]. The possibility to use a ring magnet instead of Helmholtz coils [12] is another advantage of laser frequency locking using nano-cell.

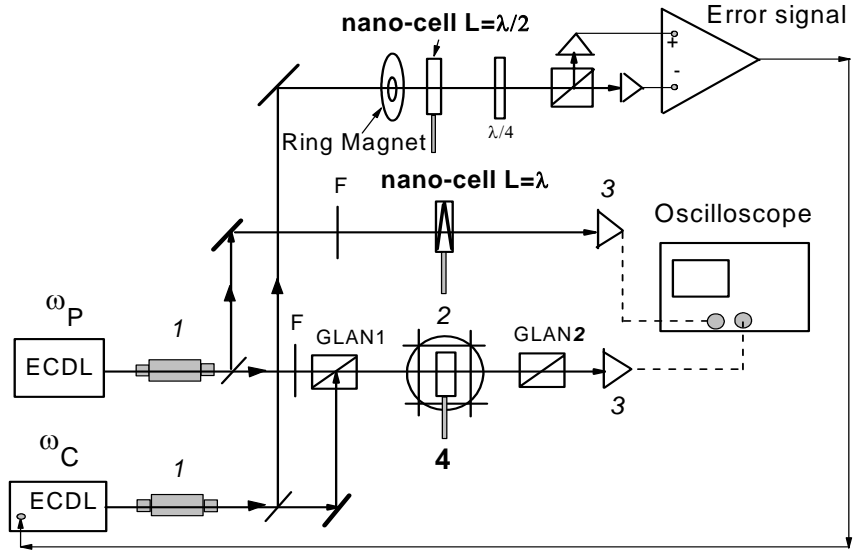

Fig.1. Sketch of the experimental setup.

Reference spectrum in the region of studied resonances was formed with the help of an auxiliary nano-cell (see Fig.1). When the nano-cell thickness is $L \approx \lambda$, sub-Doppler peaks of reduced absorption appear in the transmission spectra, centered on the hyperfine atomic transitions [16], similar to saturated absorption technique. Note that since a single beam is used, the crossover resonances that are dominant in saturated absorption spectra are absent [16], which is convenient and advantageous for frequency reference implementation. The operation temperature regime of the reference nano-cell is $T_W \sim 120^\circ C$ at the window and $T_{SA} \sim 100^\circ C$ at the side-arm.

## III. TWO-PHOTON-LIKE RESONANCE FORMATION IN THE PROBE TRANSMISSION SPECTRUM

The chosen Rb level configuration system shown in Fig.2 consists of two ground hyperfine levels of $^{85}$Rb spaced by $\Delta$ = 3036 MHz, and an excited $5P_{3/2}$ state, which serves as a common upper level. In the case when well-known EIT condition is fulfilled ($\Delta_1 = \Delta_2$), the EIT resonance is well observable in the probe transmission spectrum shown in Fig.3. For this case the frequency of the coupling laser $\nu_c$ is resonant with $5S_{1/2}$, $F_g$=2 → $5P_{3/2}$, $F_e$=3 transition, while the probe laser frequency $\nu_p$ is scanned across $5S_{1/2}$, $F_g$=3 → $5P_{3/2}$, $F_e$=2,3,4 transitions. The radiation powers of the coupling and probe lasers are ~300 μW and ~20 μW, correspondingly. The EIT resonance linewidth is ~3 MHz (this relatively large value is caused by the fact that the lasers are not coherently coupled). It is well known that buffer gas causes reduction of EIT resonance linewidth [7], notably when coherently coupled coupling and probe laser beams are used. In this case the linewidth of EIT resonance can be as narrow as ~40 ÷ 50 Hz [4]. The origin of this effect is mainly linked with essential increase of the interaction time (i.e. time of flight of atoms through the laser beam with diameter *D*).

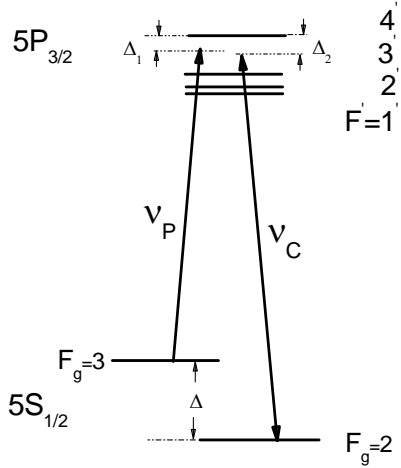
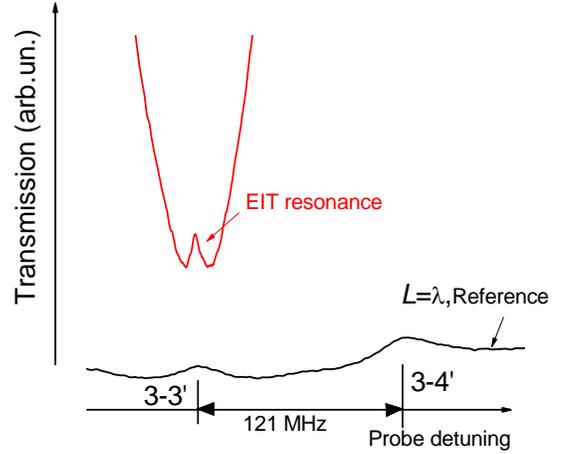

Fig.2. Λ-scheme for "dark" resonance formation.

Fig.3. Upper curve: the probe transmission spectrum with EIT-resonance; lower curve: the reference spectrum.

If now the coupling laser frequency $\nu_c$ is tuned down to be resonant with $F_g$=3 → $F_e$=4 transition, a EIA-resonance can be detected in the transmission spectrum of the probe laser which is scanned across $F_g$=3 → $F_e$=4 transition (not shown). With the further reduction of the coupling laser frequency $\nu_c$ and scan of the probe laser frequency $\nu_p$ across $F_g$=3 → $F_e$=2,3,4 transitions, a narrow "bright" resonance is formed again, when the $\nu_p - \nu_c$ frequency difference is equal to the ground state hyperfine splitting $\Delta$ (see

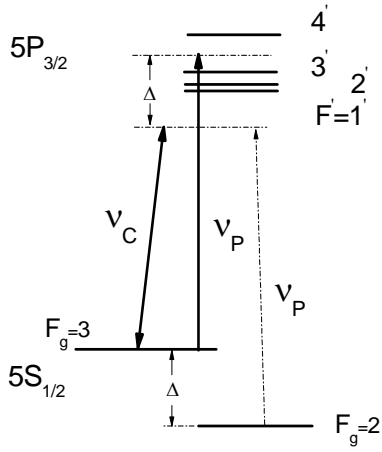
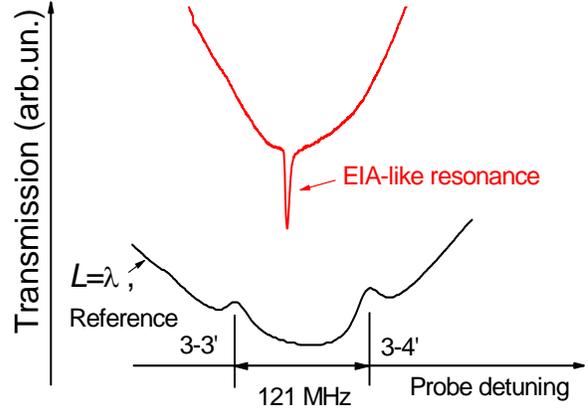

Fig.4. Configuration of $\nu_c$ and $\nu_p$ for "bright" resonance formation.

Fig.5. ASC with Rb and Ne buffer gas. Upper curve: the probe transmission spectrum with "bright" resonance; lower curve: reference spectrum.

Fig.4). This bright "EIA-like" resonance exhibits an increase of absorption in the transmission spectrum (upper curve in Fig.5) and is visually similar to EIA resonance. Meanwhile the configuration and condition of appearance differ from that needed for "genuine" EIA resonance formation [6]. Powers of couple and probe lasers for this case (3 mW and 1 mW, correspondingly) are larger by 2 orders than those needed for EIT resonance formation; the temperature of the ASC with Rb and 6 Torr of Ne buffer is $T_W \sim 80^{\circ}$C at the window and $T_{SA} \sim 60^{\circ}$C at the side-arm. The linewidth of the "EIA-like" resonance is $\sim 4.5$ MHz, i.e. less than natural linewidth of Rb ($\sim 6$MHz). The lower curve in Fig.5 is the reference spectrum of the nano-cell with the thickness $L \approx \lambda$. Note that tuning of the coupling laser frequency results in tuning of the frequency of this "EIA-like" bright resonance.

The bright resonance is seen also in fluorescence spectrum shown in Fig.6 (upper curve). The position of this increased fluorescence resonance with $\sim 12$ MHz linewidth is indicated by an arrow. Middle curve is the transmission spectrum of the probe laser (also here the bright resonance is marked by an arrow). The radiation powers are $P_C = 9$ mW, $P_P = 1$ mW, beam $\varnothing = 2$ mm. The lower curve is the reference spectrum of the nano-cell with $L \approx \lambda$ thickness.

Increase of ASC temperature to $T_W \sim 100^{\circ}$C and $T_{SA} \sim 80^{\circ}$C, increase of the coupling and probe radiations powers to $\sim 12$ mW, and also focusing radiations inside the ASC by $F = 30$ cm lens to a spot of $\sim 0.8$ mm result in increase of the bright resonance contrast and simultaneous power broadening of its linewidth to 29 MHz, as shown in Fig.7. The configuration of $\nu_c$ and $\nu_p$ frequencies for this case is shown in Fig.8. Note that the probe laser transmission is zero for low probe power (~10 μW), while for high probe intensity the transmission increases up to ~70% due to strong optical pumping effect.

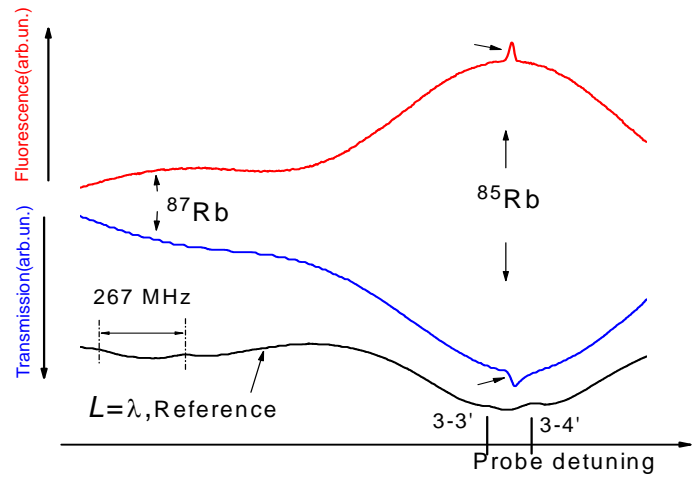

Fig.6. ASC filled with the Rb and Ne buffer gas. Upper and middle curves are the fluorescence and probe transmission spectra, correspondingly; lower curve is the reference spectrum.

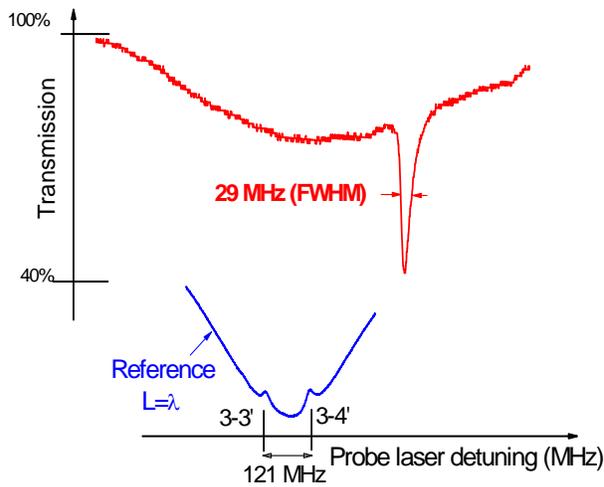
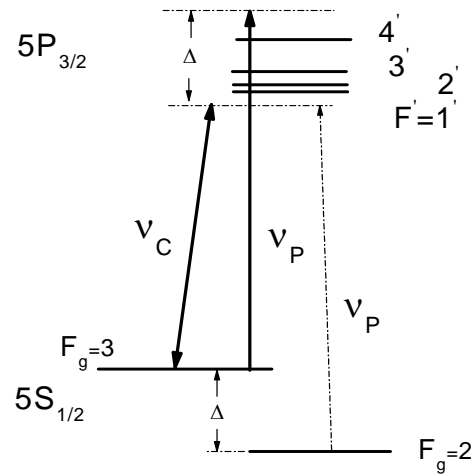

Fig.7. ASC filled with the Rb and Ne buffer gas. Upper curve: probe transmission spectrum with "bright" resonance; lower curve: the reference spectrum.

Fig.8. Configuration of $\nu_c$ and $\nu_p$ for "bright" resonance formation.

In [17], a narrow coherent resonance, which demonstrates sign reversal from "dark" to "bright", has been formed in the Rb cell with buffer gas; however the configuration of the coupling and probe laser frequencies was different from that presented in Fig.9. Authors of [17] note that the resonance they have observed in a buffered cell was absent in a cell filled with pure Rb (without buffer gas). In contrast, the bright resonance reported in this paper is still observable in a 0.7 mm-thin cell with pure Rb with $T_W \sim$ 120°C and $T_{SA} \sim$ 100°C (i.e., in the regime of relatively strong Rb–Rb collisions). The spectra for the case of $L$=0.7 mm Rb cell without buffer gas are shown in Fig.9 for $P_C$ = 11 mW and $P_P$ = 1 mW (polarizations of the coupling and probe lasers are linear and perpendicular to each other). The upper curve is transmission spectrum when the coupling laser is off. Also broad dips of decreased absorption located at the frequency position of the crossover resonances are observable in the spectra. They appear because of reflection of the probe laser radiation from the inner surface of 0.7 mm-thin Rb cell. Indeed, the cell sapphire window with the refractive index $n \approx 1.8$ forms a strong (8%) counter-propagating beam thus causing saturated absorption-like spectrum for the transmitted probe beam. The middle curve in Fig.9 is the transmission spectrum of the probe laser when the coupling laser is on. As seen, a peak of increased absorption (bright resonance) appears also in this case, though with a poor contrast as compared with the case of buffered Rb cell (Figs.5 and 7). The lower curve in Fig.9 is the reference spectrum of the nano-cell with the thickness $L \approx \lambda$.

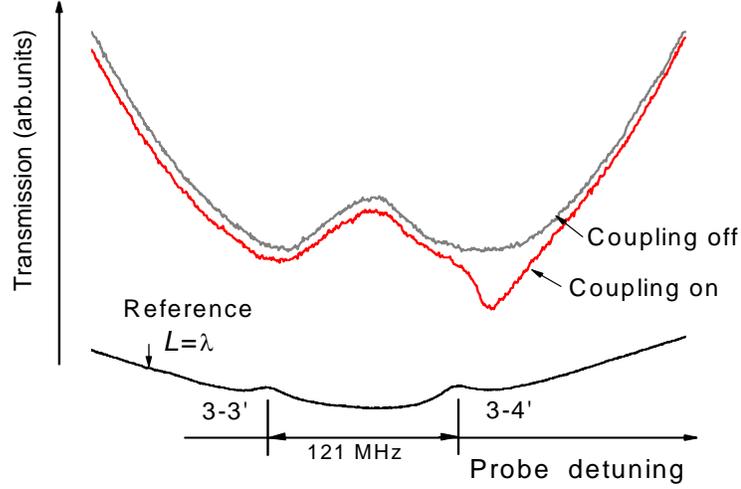

Fig.9. Cell filled with pure Rb (without buffer gas), $L$ = 0.7 mm. The upper and middle curves are the probe transmission spectra when the coupling is on and off, correspondingly; lower curve: the reference spectrum.

We believe the physical origin of formation of this "bright" resonance is as follows. The probe laser radiation with a relatively high power causes a strong optical pumping effect, which transfers a large number of Rb atoms from the ground level F=3 to the ground level F=2 (see the diagrams in Fig.4 and Fig.8). The presence of buffer gas in the ASC with Rb enhances optical pumping process. Indeed, the efficiency of optical pumping $\eta$ is proportional to the product $\Omega_L \times t$, where $\Omega_L$ is the Rabi frequency of the laser radiation and $t$ is the interaction time of atom with the laser beam. In the case of a buffered cell, the interaction time increases by several orders of magnitude compared with a cell with pure Rb. This causes efficient population transfer from the ground level F=3 to F=2, thus assuring condition $N_2 > N_3$. For $\nu_p - \nu_c = \Delta$, a strong 2-photon absorption (TPHA) of the probe radiation $5S_{1/2}$, $F_g=2 \to 5S_{1/2}$, $F_g=3$ can occur through the intermediate state $5P_{3/2}$ at the frequency $\nu_p = \nu_c + \Delta$, resulting in formation of resonance of increased absorption. The cross-section for 2-photon absorption of the probe beam $\nu_p$ is

$$\sigma_{2PHA} = \frac{\lambda^2}{16\pi^2} \cdot \frac{\gamma_N}{\gamma_{21}} \cdot \left|\frac{d \cdot E_c}{\hbar \cdot \Delta}\right|^2 , \qquad (1)$$

where $E_C$ is the electric field of the coupling laser radiation (note that the cross-section of probe laser TPHA depends on intensity of only the coupling laser), $d$ is the matrix element of the atomic transition dipole moment, $\Delta$ is the laser frequency detuning from $5P_{3/2}$ level ($\Delta \sim 3$ GHz), $\gamma_N = 6$ MHz, $\gamma_{21} \sim 0.1$ MHz, $\lambda = 780$ nm, $[(d \times E_C)/\hbar] = \Omega_C \approx 60$ MHz.

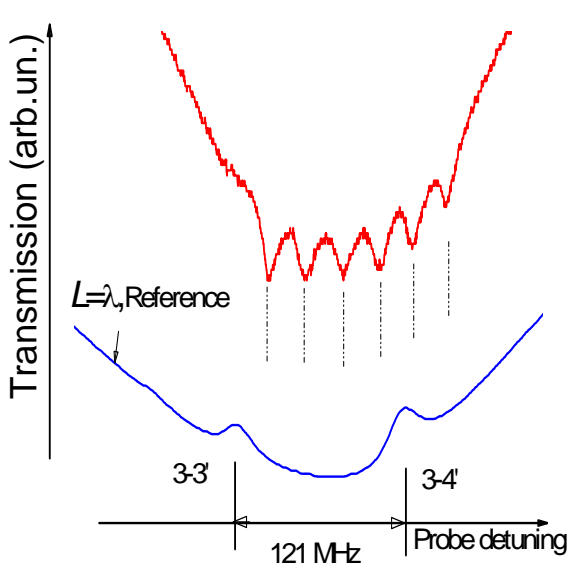 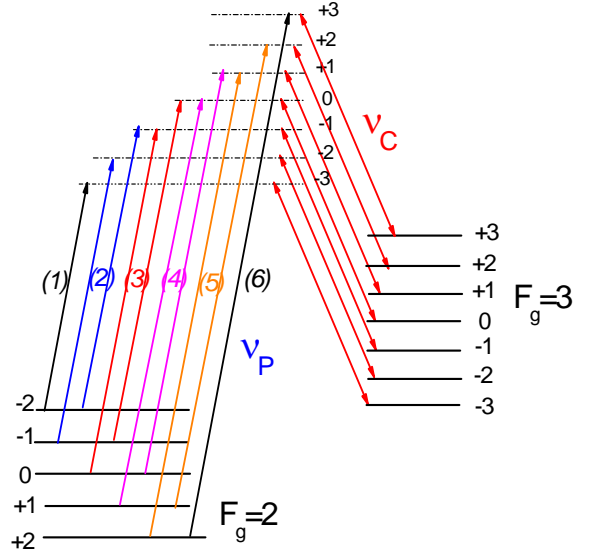

Fig.10. ASC filled with the Rb and Ne buffer gas. Upper curve: probe transmission spectrum with six TPHA-like resonances; lower curve: the reference spectrum.

Fig.11. Six pairs of the coupling $\nu_c$ and probe $\nu_p$ frequencies forming six TPHA-like resonances.

For TPHA of the probe laser one has $\sigma_{TPHA} \times N_{Rb} \times L$, where $N_{Rb} \sim 10^{11}$ cm$^{-3}$, $L \sim 0.8$ cm, and TPHA is $\sim 10\%$. Note that triple reduction of the coupling laser intensity causes disappearance of the bright resonance. It is important to note that as it is seen from Figs.5-7, the shape of TPHA resonance is not symmetrical, demonstrating some absorption reduction at the low-frequency wing. We suppose there is some additional competing process, probably EIT (a similar situation is described in [17]), which influences the TPHA resonance. That is why we prefer to consider this resonance as TPHA-like one.

We have studied the behavior of the TPHA-like resonance in an external *B*-field. The transverse magnetic field $B$ = 30 G was applied to the ASC with the help of Helmholtz coils along the coupling field *E*. The powers of coupling and probe radiations were 9 mW and 1 mW, respectively. In Fig.10 it is shown the splitting of the TPHA-like resonance into six components. Due to Zeeman effect, the splitting of atomic levels in external magnetic field occurs. The corresponding diagram is presented in Fig.11. Since polarizations of coupling and probe lasers are linear and perpendicular to each other, the probe can be considered as combination of right- and left-circular polarized radiations causing transitions shown in Fig.11. One can see that the probe laser frequency for some transitions coincides, and the total number of pairs with different probe frequencies is equal to 6. The splitting of TPHA-like resonance has been well detected for an external magnetic filed varying in the range of 10 ÷ 250 G (some increase in contrast of the resonance components is seen for $B > 150$ G).

## IV. CONCLUSION

It is demonstrated that a narrow and high-contrast "bright" resonance can be formed in the transmission spectrum of the probe laser in a coupling and probe configuration realized by two ~ 1 MHz-linewidth cw lasers with wavelengths $\lambda \approx 780$ nm in a cell filled with Rb and Ne buffer gas (pressure 6 Torr). Unlike the conditions for formation of "genuine" EIT or EIA resonances, in the proposed scheme the coupling laser frequency $\nu_c$ is stabilized and detuned by ~3 GHz from $5S_{1/2}$, $F_g$ =3 → $5P_{3/2}$, $F_e$=2,3,4 transition group, while the probe laser frequency is scanned across $5S_{1/2}$, $F_g$=3 → $5P_{3/2}$, $F_e$=2,3,4 transitions, forming a mixture of Λ- and V-systems. Narrow, contrast resonance exhibiting absorption increase and having sub-natural linewidth is detected in probe laser radiation transmission spectrum. We suppose the physical mechanism responsible for this "bright" resonance formation, which is visually similar to EIA-resonance, is the 2-photon absorption (TPHA) of the probe radiation $5S_{1/2}$, $F_g$=2 → $5S_{1/2}$, $F_g$=3 with the intermediate state $5P_{3/2}$, at the frequency $\nu_p = \nu_c + \Delta$. On the other hand, there is also a competing process taking place in the same conditions (probably, EIT), which influences the lineshape of this TPHA resonance.

It is shown that the TPHA-like resonance can be formed also in a thin (0.7 mm-long) cell filled with pure Rb (without buffer gas), however in this case a higher temperature is needed (~100°C) to produce relatively strong Rb-Rb collisions regime. Contrast of the resonance in this case is somewhat worse compared with the case of buffered cell. The TPHA-like resonance is split in an external magnetic field $B$ = 10÷250 G into 6 well resolved and high-contrast components. Physical explanation of this splitting is presented, too.

*Acknowledgements*: This work was partially carried out in the framework of the INTAS South-Caucasus Grant 06-1000017-9001.